\documentclass[12pt]{article}
\usepackage{amssymb,amsmath,epsfig}

\begin{document}

\title{\bf Dynamical Analysis of Charged Anisotropic Spherical Star in $f(R)$ Gravity}

\author{H. Rizwana Kausar \thanks{rizwa\_math@yahoo.com} ${}^{(a)}$, Ifra Noureen \thanks{ifra.noureen@gmail.com} ${}^{(b)}$,
M. Umair Shahzad \thanks{m.u.shahzad@ucp.edu.pk}
${}^{(a)}$\\${}^{(a)}$ CAMS, UCP Business School,\\
University of Central Punjab, Lahore, Pakistan.\\
${}^{(b)}$ Department of
Mathematics,\\University of Management and Technology,\\
Lahore, Pakistan. }
\date{}
\maketitle

\begin{abstract}
We consider a modified gravity theory, $f(R)=R+\alpha
R^n-\frac{\mu^4}{R^m}$, in the metric formulation and analyze the
contribution of electromagnetic field on the range of dynamical
instability of a star filled with anisotropic matter. The collapse
equation is developed by applying conservation on anisotropic
matter, Maxwell source and dark source terms arising due to $f(R)$
gravity. Specific perturbation scheme is implemented and it is
observed that the inclusion of Maxwell source slows down the
collapse and makes system more stable in Newtonian regime. Also, we
make comparison of our results with the existing literature.
\end{abstract}

\section{Introduction}

The investigations about the final outcome of a stellar collapse and
stability of a self-gravitating body is always a key issue in
astrophysics and gravitational theories. The astrophysical bodies
remain stable until the outward pressure produced by internal fusion
balances the inward acting gravity force. The endless collapsing
phenomenon takes place when a massive star exhaust its nuclear fuel,
the gravitational force then take over implying the death of a star
\cite{1}. The life cycle of a massive star having mass of the order
$10-20$ solar mass is nominal in comparison to the stars with
relatively less mass i.e., $\approx1$ solar mass. During collapse,
the energy dissipation in the form of heat flow produces different
scenarios for different sizes of stars \cite{2}-\cite{2''}. A more
massive star tends to be more unstable due to high radiation
transport rather than a less massive star.

The self-gravitating object is worthwhile only if it is stable
against fluctuations. The pioneer contribution on the problem of
dynamical instability is by Chandrasekhar \cite{3}, who found
instability range in the form of adiabatic index $\Gamma$ as
$\Gamma\geq\frac{4}{3}+n\frac{M}{r}$, where $M$ and $R$ are the mass
and radius of a star respectively. Hillebrandt and K.O. Steinmetz
\cite{5} establish the stability criterion for the model of
anisotropic neutron stars.

Herrera and his contemporaries established the stability ranges for
various forms of gravitational collapse by considering isotropy,
local anisotropy, expansion free condition, dissipation etc.
\cite{6}-\cite{8}. It was established that minor changes in fluid
isotropy affects subsequent evolution considerably, also, in
relativistic annexes the dissipative, radiative and shearing effects
gave rise to the stability of compact object \cite{9}-\cite{13} by
slowing down the collapse.

The collapse and stability of gravitating bodies in alternative
gravity theories have been studied extensively. Among modified
theories, $f(R)$ represents the most simple alteration in
Einstein-Hilbert (EH) action to include higher order curvature
invariants by replacing Ricci scalar $R$ with the terms of order
$\sqrt{-g}f(R)$. Here $f(R)$ denote the general function of $R$ and
$g$ stands for the metric tensor \cite{15}-\cite{17}. Thus the
action for $f(R)$ gravity can be written as
\begin{equation}\setcounter{equation}{1}\nonumber
S=\frac{1}{2\kappa}\int d^{4}x\sqrt{-g}f(R),
\end{equation}
where $\kappa$ is the coupling constant. The viability and stability analysis in
context of modified theories of gravity has been presented in
\cite{18}-\cite{z5} in view of observational situations such as
cosmic microwave background, clustering spectrum and weak lensing
\cite{27}-\cite{30}.

The electromagnetic field affects the evolution and structure
stability range of the collapsing relativistic bodies. The magnetic
and Coulomb forces are the major source for resisting the
gravitational pull \cite{31}. A star needs a large amount of charge
in order to stay stable among the strong gravitational pull
\cite{MA}. Such electromagnetic tension also leads to homogeneity on
large scales. In modified $f(R)$ theory, some work has been done on
the aspect of instability of collapsing stars by considering
combined effect of electromagnetic field and $f(R)$ model
\cite{MA}-\cite{35}. Motivating by this work, we have chosen
$f(R)=R+\alpha R^n-\frac{\mu^4}{R^m}$ and charged anisotropic fluid
to workout the instability problem. The matching of stable stellar
configuration corresponds to positive real values of $\alpha$, $m$
and $n$, while $\mu$ can be any real number. However, to meet
viability criteria of the proposed model, we need to fix $n\geq 2, m
\geq 1$ to keep $\frac{d^2f}{dR^2}$ positive.

The manuscript is arranged in the following manner: Section
\textbf{2} provides energy-momentum tensor for usual matter, Maxwell
source and dark source along with Einstein's field equations and
dynamical equations. In section \textbf{3}, the perturbation scheme
is presented and discussions on instability in terms of adiabatic
index is given. Summary of the obtained results is provided in
section \textbf{4} which is followed by an appendix.

\section{Evolution Equations}

Here, a three dimensional timelike spherical boundary $\Sigma$ is
chosen that imparts space time into two regions referred as interior
and exterior regions respectively. The line element for region
inside the spherical boundary surface $\Sigma$ is given by
\begin{equation}\setcounter{equation}{1}\label{1}
ds^2_-=A^2(t,r)dt^{2}-B^2(t,r)dr^{2}-C^2(t,r)(d\theta^{2}+\sin^{2}\theta
d\phi^{2}).
\end{equation}
The exterior region is defined by the line element as follow
\cite{35}
\begin{equation}\label{25}
ds^2_+=\left(1-\frac{2M}{r}+\frac{Q^2}{r^2}\right)d\nu^2+2drd\nu-r^2(d\theta^2+\sin^{2}\theta
d\phi^{2}),
\end{equation}
where $Q$ stands for the total charge, $\nu$ is the retarded time
and $M$ denotes the total mass. The EH action in $f(R)$ theory
including Maxwell source is defined as
\begin{equation}\label{b}
S=\frac{1}{2}\int
d^{4}x\sqrt{-g}\left(\frac{f(R)}{\kappa}-\frac{\digamma}{2\pi}\right),
\end{equation}
where $\digamma=\frac{1}{4}F^{uv}F_{uv}$ is the Maxwell invariant.
By varying above action with respect to the metric tensor $g_{uv}$,
we obtain the following set of field equations
\begin{equation}\label{b'}
f_RR_{uv}-\frac{1}{2}f(R)g_{uv}-\nabla_{u} \nabla_{v}f_R+ g_{uv}
\Box f_R=\kappa (T_{uv}+E_{uv}),\quad(u, v =0,1,2,3).
\end{equation}
Here $f_R\equiv df(R)/dR$, $\Box=\nabla^{u}\nabla_{v}$ where
$\nabla_{u}$ represents the covariant derivative, $E_{uv}$ is the
electromagnetic tensor and $T_{uv}$ is the minimally coupled
stress-energy tensor. Rewriting, above set of field equations in the
form of Einstein tensor on the left hand side, we have
\begin{equation}\label{12}
G_{uv}=\frac{\kappa}{f_R}[L_{uv}],
~~\textmd{where}~~L_{uv}=\overset{(D)}{T_{uv}}+T_{uv}+E_{uv}.
\end{equation}
For interior spherically symmetric metric, the components of
Einstein tensor can be found in \cite{35}. In the above equation
$\overset{(D)}{T_{uv}}$ is supposed to known as dark source energy
contribution due to modification in gravity. It is given by
\begin{equation}\label{d}
\overset{(D)}{T_{uv}}=\frac{1}{\kappa}\left[\frac{f(R)-Rf_R}{2}g_{uv}+\nabla_{u}
\nabla_{v}f_R -g_{uv} \Box f_R\right],
\end{equation}
The energy momentum tensor $T_{uv}$ is referred to the usual matter
under consideration which is assumed to be charged anisotropic fluid
dissipating energy via heat flow $q$. It is defined as
\begin{equation}\label{2}
T_{uv}=(\mu+p_\perp)V_{u}V_{v}-p_\perp
g_{uv}+(p_r-p_\perp)\chi_u\chi_v+ q_u V_v+q_{v}V_u,
\end{equation}
where $V_{u}$ is the for four-velocity, $\mu$ is the density, $p_r$
and $p_\perp$ corresponds to redial and tangential pressures
respectively and $\chi_u$ is the corresponding radial four vector.
These quantities satisfy the following relations in co-moving
coordinates
\begin{equation}\label{3}
V^{u}=A^{-1}\delta^{u}_{0},\quad q^{u}=qB^{-1}\delta^u_1,\quad
\chi^{u}\chi_{u}=-1, \quad \chi^{u}=B^{-1}\delta^{u}_{1}.
\end{equation}
The energy momentum tensor corresponding to Maxwell source is given
by \cite{35}
\begin{equation}\label{3'}
E_{uv}=\frac{1}{4\pi}(-F^w_{u}F_{vw}+\frac{1}{4}F^{wx}F_{wx}g_{uv}),
\end{equation}
where electromagnetic field tensor $F_{uv}$ is defined by
$F_{uv}=\varphi_{v,u}-\varphi_{u,v}$ with
$\varphi_{u}=\varphi(t,r)\delta^{0}_{u}$ representing the four
potential. The Maxwell's equations are \cite{36}
\begin{equation}\label{3''}
F^{uv}_{;v}=4\pi\jmath^u, \quad F_{uv;w}=0,
\end{equation}
where four current is $\jmath^u=\xi(t,r)V^u$ and $\xi$ stands for
charge density. On simplification, electromagnetic field equations
become
\begin{eqnarray}\label{M1}
&&\frac{\partial^2\varphi}{\partial
r^2}-\left(\frac{A'}{A}+\frac{B'}{B}-\frac{2C'}{C}\right)
\frac{\partial\varphi}{\partial r}=4 \pi \mu AB^2,
\\\label{M2}&&\frac{\partial^2\varphi}{\partial t \partial r}-\left(\frac{\dot{A}}{A}
+\frac{\dot{B}}{B}-\frac{2\dot{C}}{C}\right)\frac{\partial\varphi}{\partial
r}=0.
\end{eqnarray}
Herein dot and prime indicate time and radial derivatives
respectively. The charge conservation is defined as
\begin{equation}\label{3'''}
s(r)={\int}^{r}_{0}\mu BC^2 dr,\quad E=\frac{s}{4\pi C^2},
\end{equation}
where time independent charge interior to radius $r$ is denoted by
$s$ and $E$ is the electric field intensity. Integrating
Eq.(\ref{M1}), we get
\begin{equation}\label{3'''}
\frac{\partial\varphi}{\partial r}=\frac{sBA}{C^2}.
\end{equation}

The matching conditions across boundary surface are of great interest in order to seek the valid solution to the field equations for both the interior and exterior metrics. The smooth matching conditions corresponding to Darmois junction conditions \cite{38} imply that the induced metric and extrinsic curvature must be continuous on both sides of the hypersurface. Relativistic objects bearing discontinuity in the extrinsic curvatures can be
described by the presence of a thin-shell formalism presented by Israel \cite{39, 39'}.
Matching of interior and exterior spacetimes across $\Sigma$ yield
\begin{equation}\nonumber
m(t,r)\overset\Sigma=M \Longleftrightarrow q(r)\overset\Sigma=Q, \quad p_\perp\overset\Sigma=\frac{-1}{B^2}\left(\frac{B}{A}\overset{(D)}{T_{01}}+\overset{(D)}{T_{11}}\right),
\end{equation}
where $m(t, r)$ denotes the Misner-Sharp mass given by \cite{40}
\begin{equation}\nonumber
m(t,r)=\frac{C}{2}\left(1+\frac{\dot{C}^2}{A^2}-\frac{C'^2}{B^2}\right)+\frac{q^2}{2C}.
\end{equation}

The Bianchi identities/dynamical equations are employed to workout
the evolution of gravitating objects by applying conservation on
usual matter along with dark source terms and electromagnetic energy
momentum tensor as
\begin{eqnarray}\label{bb}
L^{uv}_{;v}V_{u}=0,\quad L^{uv}_{;v}\chi_{u}=0.
\end{eqnarray}
Consequently, we obtain following two equations
\begin{eqnarray}\nonumber &&\dot{\rho}+q'\frac{A}{B}+2q\frac{A}{B}\left(\frac{A'}{A}
+\frac{C'}{C}\right)+\rho\left(\frac{\dot{B}}{B}+\frac{\dot{C}}{C}\right)+
p_r\frac{\dot{B}}{B}+2p_\perp \frac{\dot{C}}{C}\\\label{B1}&&+
P_1(r,t)=0,\\\nonumber &&
p'_r+p_r\left(\frac{A'}{A}+\frac{C'}{C}\right)+\rho\frac{A'}{A}-2p_\perp\frac{C'}{C}+\dot{q}\frac{B}{A}+2\frac{B}{A}\left(\frac{\dot{B}}{B}
+\frac{\dot{C}}{C}\right)\\\label{B2}&&-4\pi
E^2\left(\frac{E'}{E}+\frac{2C'}{C}\right)+P_2(r,t)=0.
\end{eqnarray}
The dark matter terms in $P_1(r,t)$ and $P_2(r,t)$ are given in
\textbf{Appendix A} as Eqs.(\ref{B3}) and (\ref{B4}) respectively.

\section{The Perturbed Dynamical Equations}

The perturbation scheme is being employed here to acquire the
simplified form of non-linear field equations whose general solution
has not been established yet. The perturbed form of dynamical
equations lead to the linear equations in the form of material and
metric variables. The dynamics of evolution can be investigated
either by following Eulerian or Lagrangian pattern, i.e., fixed or
co-moving coordinates respectively. Here we have chosen co-moving
coordinates in which it is assumed that all the quantities are in
static equilibrium initially. However, with the passage of time
these quantities have small perturbations in the form of $\epsilon$
and acquire time dependence too. In this approach, our chosen $f(R)$
model which is
\begin{equation}\setcounter{equation}{1}\label{fr}
f(R)=R+\alpha R^n-\frac{\mu^4}{R^m}
\end{equation}
would also be perturbed along the same pattern. Thus, we can write
all the quantities in their perturbed form as
\begin{eqnarray}\label{41'}
A(t,r)&=&A_0(r)+\epsilon T(t)a(r),\\\label{42'}
B(t,r)&=&B_0(r)+\epsilon T(t)b(r),\\\label{43} C(t,r)&=&r+\epsilon
T(t)\bar{c}(r),\\\label{44'} \rho(t,r)&=&\rho_0(r)+\epsilon
{\bar{\rho}(t,r)},\\\label{45'} p_{r}(t,r)&=&p_{r0}(r)+\epsilon
{\bar{p}_{r}(t,r)},\\\label{46'}
p_{\perp}(t,r)&=&p_{\bot0}(r)+\epsilon
{\bar{p}_{\bot}(t,r)},\\\label{48'} q(t,r)&=&\epsilon
{\bar{q}}(t,r),\\\label{47'} E(t,r)&=&E_0(r)+\epsilon T(t)h(r),
\\\label{49'}R(t,r)&=&R_0(r)+\epsilon T(t)e(r),\\\nonumber f(R)&=&\left(R_0+\alpha
R_0^n-\mu^4 R_0^{-m}\right)+\epsilon T(t)e(r)\left(1+\alpha
nR_0^{n-1}\right.\\\label{50'}
&+&\left.\mu^4mR_0^{-m-1}\right),\\\nonumber f_R(R)&=&\left(1+\alpha
nR_0^{n-1}+\mu^4mR_0^{-m-1}\right)+\epsilon T(t)e(r)\left[\alpha
n(n-1)R_0^{n-2}\right.\\\label{51'}
&-&\left.\mu^4m(m+1)R_0^{-m-2}\right].
\end{eqnarray}
Applying above set of equations on Bianchi identities (\ref{B1}) and
(\ref{B2}), we have
\begin{eqnarray}\nonumber
&&\frac{\bar{\rho}}{A_0^2}+\frac{2\bar{q}A_0'}{A_0^2B_0}+\frac{\bar{q}'}{A_0B_0}+\frac{2\bar{q}}{rA_0B_0}
+\left[\frac{b(p_{r0}+\rho_0)}{A_0^2B_0^2}\right.\\\label{B1p}&&\left.
+\frac{2\bar{c}(p_{\bot0}-\rho_0)}{rA_0^2}+P_{1p}\right]\dot{T}=0,\\\nonumber
&&\frac{\dot{\bar{\rho}}}{A_0B_0}+\frac{\bar{p_r}'}{B_0^2}+
\frac{A_0'}{A_0B_0^2}(\bar{\rho}+\bar{p_r})+\frac{2}{rB_0^2}(\bar{p_r}-\bar{p_\bot})
+\left[\frac{A_0'}{A_0B_0^2}(\rho_{0}\right.\\\nonumber
&&\left.+p_{r0})\left(\frac{a'}{A_0'}-\frac{a}{A_0}-\frac{2b}{B_0}\right)
+\frac{2}{rB_0^2}(p_{r0}-p_{\bot
0})\left(\bar{c}'-\frac{\bar{c}}{r}-\frac{2b}{B_0}\right)\right.\\\label{B2p}&&\left.-4\pi\left\{(E_0
h )'+2E_0^2\left(\frac{\bar{c}}{r}\right)'+2
E_0h(\frac{E_0'}{E_0}+\frac{2}{r})\right\}+P_{2p}\right]T=0.
\end{eqnarray}
The perturbed configurations of $P_{1}$ and $P_{2}$ are denoted by
$P_{1p}$ and $P_{2p}$ respectively and are provided in the
\textbf{Appendix B}, Eqs.(\ref{B5})-(\ref{B6}). Heat flux can be found
by applying perturbation on the field equation $G_{01}=L_{01}$. The
obtained relation is given as follow
\begin{eqnarray}\nonumber
\bar{q}&=&\left[\frac{\alpha R_0^{n-1}}{\kappa
A_0B_0}\left\{(n-1)R_0^{-1}+(n-1)(n-2)R_0^{-2}R_0'e-(n-1)R_0^{-1}\left(\frac{eA_0'}{A_0}\right.\right.\right.\\\nonumber
&&\left.\left.\left.\left.+\frac{b}{B_0}\right)
+2\left(\frac{\bar{c}'}{r}-\frac{\bar{c}A_0'}{rA_0}-\frac{b}{rB_0}\right)\right\}
+\frac{\mu^4 m R_0^{-m-1}}{\kappa
A_0B_0}\{-(m+1)R_0^{-1}e'\right.\right.\\\nonumber
&&\left.\left.+(m+1)(m+2)R_0^{-2}R_0'e+(m+1)R_0^{-1}\left(\frac{eA_0'}{A_0}+\frac{b}{B_0}\right)+2\left(\frac{\bar{c}'}{r}
\right.\right.\right.\\\label{q}
&&\left.\left.\left.-\frac{\bar{c}A_0'}{rA_0}-\frac{b}{rB_0}\right)\right\}+\frac{2}{\kappa
A_0B_0}\left(\frac{\bar{c}'}{r}-\frac{\bar{c}A_0'}{rA_0}-\frac{b}{rB_0}\right)\right]\dot{T}.
\end{eqnarray}
By substituting above value of $\bar{q}$ and its derivative in
Eq.(\ref{B1p}), we obtain
\begin{equation}\label{rho}
\bar{\rho}=-\left[(p_{r0}+\rho_{0})\frac{b}{B_0^2}+(p_{\bot0}-\rho_{0})\frac{2\bar{c}}{r}+P_3\right]T.
\end{equation}
Here expression for $P_3$ constitutes the dark source terms of the
proposed $f(R)$ model and is given in the \textbf{Appendix B}. The
Harrison-Wheeler type equation of state describes second law of
thermodynamics, that associates $\bar{\rho}$ and $\bar{p_r}$ in the
following pattern
\begin{equation}\label{p}
\bar{p_r}=\Gamma\frac{p_{r0}}{\rho_0+p_{r0}} \bar{\rho},
\end{equation}
where $\Gamma$ is the adiabatic index that amounts the pressure
fluctuation with density variation. Insertion of Eq.(\ref{B7}) in
the above equation, it implies
\begin{equation}\label{pr}
\bar{p_r}=-\Gamma
\left[p_{r0}\frac{b}{B_0^2}+p_{r0}\left(\frac{p_{\bot 0}
-\rho_0}{p_{ro}+\rho_0}\right)\frac{2\bar{c}}{r}+\frac{p_{r0}}{p_{r0}+\rho_0}P_3\right]T.
\end{equation}
Substituting $\bar{q}$ , $\bar{\rho}$, and $\bar{p}_r$ from
Eqs.(\ref{q}), (\ref{rho}) and (\ref{pr}) respectively in
(\ref{B2p}), we obtain the following required collapse equation
\begin{eqnarray}\nonumber
&&-\Gamma
\left[p_{r0}\frac{b}{B_0^2}+p_{r0}\left(\frac{p_{\bot0}-\rho_{0}}{\rho_{0}
+p_{r0}}\right)\frac{2\bar{c}}{r}+\frac{p_{r0}}{p_{r0}+\rho_0}\left\{P_3+4\pi
E_0^2\left(\frac{h}{E_0} \right.\right.\right.\\\nonumber
&&\left.\left.\left.\left.+\frac{2\bar{c}}{r}\right)\right\}\right]'T
-\left[\frac{A_0'}{A_0}+\Gamma
\frac{p_{r0}}{\rho_0+p_{r0}}\left(\frac{A_0'}{A_0}+\frac{2}{r}\right)\right]
\left[(\rho_{0}+p_{r0})\frac{b}{B_0^2}\right.\right.\\\nonumber
&&\left.\left.+(p_{\bot0}-\rho_{0})\frac{2\bar{c}}{r}+P_3+4\pi
E_0^2\left(\frac{h}{E_0}
+\frac{2\bar{c}}{r}\right)\right]T-\frac{2}{r}p_{\bot}+\left[\frac{-2b
p_{ro}'}{B_0}\right.\right.\\\nonumber
&&\left.\left.+\frac{A_0'}{A_0}(\rho_{0}+p_{r0})
\left(\frac{a'}{A_0'}-\frac{a}{A_0}-\frac{2b}{B_0}\right)+\frac{2}{r}(p_{r0}
-p_{\bot0})\left(\bar{c}-\frac{\bar{c}}{e}-\frac{2b}{B_0}\right)\right.\right.\\\nonumber
&&\left.-4\pi\left\{(E_0 h )'+2E_0^2(\frac{\bar{c}}{r})'+2
E_0h(\frac{E_0'}{E_0}+\frac{2}{r})\right\} +P_{2p}\right]T+
\left[\frac{\alpha n R_0^{n-1}}{\kappa A_0^2}\right.\\\nonumber
&&\times\left.\left.\left[(n-1)R_0^{-1}e'+(n-1)(n-2)R_0^{-2}R_0'e-(n-1)R_0'\left(e\frac{A_0'}{A_0}+\frac{b}{B_0}\right)
\right.\right.\right.\\\nonumber &&\left.\left.\left.+
2\left(\frac{\bar{c}'}{r}-\frac{\bar{c}A_0'}{rA_0}-\frac{b}{rB_0}\right)\right]
+\frac{\mu^4 m R_0^{-m-1}}{\kappa
A_0B_0}\left[-(m+1)R_0^{-1}e'+(m\right.\right.\right.\\\nonumber
&&\left.+1)(m+2)R_0^{-2}R_0'e+(m+1)R_0^{-1}\left(\frac{eA_0'}{A_0}
+\frac{b}{B_0}\right)+2\left(\frac{\bar{c}'}{r}-\frac{\bar{c}A_0'}{rA_0}\right.\right.\\\label{G1}
&&\left.\left.\left. -\frac{b}{rB_0}\right)\right]+\frac{2}{\kappa
A_0B_0}\left(\frac{\bar{c}'}{r}-\frac{\bar{c}A_0'}{rA_0}-\frac{b}{rB_0}\right)\right]\ddot{T}=0.
\end{eqnarray}
This equations helps to study the evolution of spherical charged
collapsing star. The perturbed configuration of Ricci scalar yields
an ordinary differential equation in the form below
\begin{equation}\label{66}
\ddot{T}(t)-P_4(r) T(t)=0,
\end{equation}
where $P_4$ is provided in the \textbf{Appendix B}, Eq.(\ref{B9}).  In
order to establish instability range, the terms appearing in $P_4$
are assumed in such a way that over all expression should be
positive. The solution of Eq.(\ref{66}) is given by
\begin{equation}\label{68}
T(t)=-e^{\sqrt{P_4}t}.
\end{equation}

\section*{Newtonian Limit}

To approximate our results in Newtonian gravity, we assume and
substitute $\rho_0\gg p_{r0}$, $\rho_0\gg p_{\perp0}$ and
$A_0=1,~B_0=1$ in Eq.(\ref{G1}) and get collapse equation as
\begin{eqnarray}\nonumber
&&\left[\Gamma
\left(-p_{r0}b\right)'T-\frac{2}{r}p_{\bot}\left(\frac{1}{T}\right)-2b
p_{ro}'+a'\rho_{0}+\frac{2}{r}(p_{r0}-p_{\bot0})\left(\bar{c}'-\frac{\bar{c}}{r}-2b\right)
\right.\\\nonumber &&\left.+P_{2P}^{N}\right]T+\left[\frac{\alpha n
R_0^{n-1}}{\kappa}\left[(n-1)R_0^{-1}e'+(n-1)(n-2)R_0^{-2}R_0'e-(n-1)R_0'b
\right.\right.\\\nonumber
&&\left.\left.+2\left(\frac{\bar{c}'}{r}-\frac{b}{r}\right)\right]+\frac{\mu^4
m R_0^{-m-1}}{\kappa }\left[-(m+1)R_0^{-1}e'+(m+1)(m+2)R_0^{-2}R_0'e
\right.\right.\\\nonumber
&&\left.\left.+(m+1)R_0^{-1}b\right]\frac{2}{\kappa
}\left(\frac{\bar{c}'}{r}-\frac{b}{r}\right)\right]\ddot{T}-4\pi\left\{(E_0
h )'+2E_0^2(\frac{\bar{c}}{r})'+2
E_0h(\frac{E_0'}{E_0}\right.\\\label{53}&&
\left.+\frac{2}{r})\right\}=0.
\end{eqnarray}
Here, $P_{2p}^{N}$ represents the Newtonian approximation of
perturbed second Bianchi identity. Substituting value of $T$ from
Eq.(\ref{68}) and its subsequent time derivatives in the above
equation, we get
\begin{equation}
\Gamma < \frac{a'\rho_{0}-2b
p_{ro}'+\frac{2}{r}(p_{r0}-p_{\bot0})\left(\bar{c}'-\frac{\bar{c}}{r}-2b\right)+E_1+P_{2p}^{N}+P_5(r)}{\left(-p_{r0}b\right)'}
\end{equation}
where $E_1=-4\pi\left\{(E_0 h )'+2E_0^2(\frac{\bar{c}}{r})'
+2E_0h(\frac{E_0'}{E_0}+\frac{2}{r})\right\}$ and $P_5$ is given in
the \textbf{Appendix B}. The terms involving in $E_1$ depicts that the
resistance offered to the inward drawn gravity include
electromagnetic field components detaining collapse and hence imply
a wider range of stability. The expression $\Gamma$ have various
limiting cases in the Newtonian regime as follows:
\begin{enumerate}
\item when $\mu \rightarrow 0$, the expression for $\Gamma$ will remains
the same for usual matter, however, effective part $P_5$ reduced to
$P_6$ as
\begin{eqnarray}\nonumber
P_6(r)&=&P_4\left[\frac{\alpha n R_0^{n-1}}{\kappa
}\left[(n-1)R_0^{-1}e'+(n-1)(n-2)R_0^{-2}R_0'e-(n\right.\right.\\\label{P6}
&&\left.\left.-1)R_0'b+2\left(\frac{\bar{c}'}{r}-\frac{b}{r}\right)\right]+\frac{2}{\kappa
}\left(\frac{\bar{c}'}{r}-\frac{b}{r}\right)\right].
\end{eqnarray}
\item  when $\alpha\rightarrow 0$, $P_4$ changes to $P_7$, whereas terms involving matter
variables remain unchanged. The expression for $P_7$ is given by
\begin{eqnarray}\nonumber
P_7(r)&=&P_4\left[\frac{\mu^4 m R_0^{-m-1}}{\kappa
}\left[-(m+1)R_0^{-1}e'+(m+1)(m+2)R_0^{-2}R_0'e\right.\right.\\\label{P7}
&&\left.\left.+(m+1)R_0^{-1}b\right]+\frac{2}{\kappa
}\left(\frac{\bar{c}'}{r}-\frac{b}{r}\right)\right].
\end{eqnarray}
\item When both $\alpha\rightarrow 0$ and $\mu \rightarrow
0$, $\Gamma$ becomes
\begin{equation}
\Gamma < \frac{a'\rho_{0}-2b
p_{ro}'+\frac{2}{r}(p_{r0}-p_{\bot0}+\frac{p_{\bot}}{r})\left(\bar{c}'-\frac{\bar{c}}{r}-2b\right)
+E_1+\frac{P_{4}(r)}{\kappa}\left(\frac{2\bar{c}'}{r}
-\frac{2b}{r}\right)}{\left(-p_{r0}b\right)'}
\end{equation}
representing the GR case.
\end{enumerate}
It is mentioned that the above mentioned limiting cases are
consistent with the previously done work on $f(R)$ models. Also, for
the post Newtonian limit, the relativistic effects can be analyzed
upto $O(\frac{m_0}{r}+ \frac{Q^2}{2r^2})$ by assuming metric
coefficients as
\begin{eqnarray}\label{pn1}
&&A_0=1-\frac{m_0}{r}+\frac{Q^2}{2r^2},~~~~B_0=1+\frac{m_0}{r}-\frac{Q^2}{2r^2}, \\\label{pn2}
\Rightarrow~&&\frac{A_0'}{A_0}=\frac{2}{r}\frac{Q^2-rm_0}{2rm_0-2r^2-Q^2},~~~\frac{B_0'}{B_0}
=\frac{2}{r}\frac{Q^2-rm_0}{2rm_0+2r^2-Q^2}.
\end{eqnarray}
In this case adiabatic index $\Gamma$ would have much dependence on
the electric field configuration, pressure and density terms.

\section{Summary}

The present work is devoted to investigate the consequences of
$f(R)$ model and electromagnetic field surrounding the spherically
symmetric stars. The model $f(R)=R+\alpha R^n-\frac{\mu^4}{R^m}$ is
a viable framework to assume in this dark energy era. The usual
matter configuration of star is charged anisotropic fluid which
dissipate energy in terms of radiation streaming and heat flow and
hence effect stability drastically. As modified field equations are
highly complicated non-linear differential equations whose solution
has not been established yet. Thus, we have applied perturbation
approach, on the dynamical equations to observe evolution of
different variables at the surface of a collapsing star.

It is found that instability range of assumed star depend upon the
adiabatic index $\Gamma$ which further depends upon the tangential
and radial pressure, density, electric field and radiation etc. It
is worthwhile to mention that the electromagnetic effects along with
the higher order curvature terms contributions builds up the
stability range because dissipative effects decelerate the collapse.
The observations for limiting cases $\alpha\rightarrow0,
\beta\rightarrow0$ reduce the results to GR.

The strong gravitational field impressions can be anticipated by analyzing the dark source entries appearing in collapse equation (3.19), most of them appears to be negative in equation implying instabilities. We may argue that strong gravitational field limit bring instabilities more rapidly in comparison to the weak field limit implying gravitational collapse.
Furthermore, it is worth
stressing that the numerical pattern devised in \cite{41}-\cite{43} can
be taken into account to address instability problem in an
alternate and adequate way. The strong field limit can be established more appropriately by constraining to some specific form of relativistic object and present Jeans analysis.

By comparing our work with the literature, it is observed that by
vanishing of parameters involved in the chosen $f(R)$, the obtained
reduced expressions for $\Gamma$ yield consistency with the already
work done for the independent $f(R)$ models. The comparison is as
follows
\begin{itemize}
\item In \cite{32}, the effects of Maxwell source and CDTT model on dynamical instability
 of locally isotropic background were discussed, the results can be retrieved for our consideration by assuming $p_r=p_\perp, m=-1,
  \mu=\delta, m=1$ and $\alpha=0$.
\item  For $n=2, \alpha=\delta$ and $q_\alpha=0$ in our assumptions the findings are in accordance with the
work done in \cite{34} for $f(R)=R+\delta R^2$ model on dynamical
instability.
 \item When we take $\mu=0$, the results support the arguments in \cite{35}. Also, it is obvious
  that addition of Maxwell invariant describe more general
 expanding universe with a wider range of instability in $f(R)$ framework.
 \item The work on generalized CDTT model in the absence of Maxwell source appears to be special case of our model
by assuming $ E=0, Q=0, p_r=p_\bot, n=2$ and $m=1$. The results for
these substitutions in our considerations are in agreement with the
findings of a recent paper \cite{37}.
\end{itemize}

\section*{Appendix A: Unperturbed Quantities}

\begin{eqnarray}\setcounter{equation}{1}\nonumber
P_1(r,t)&=&\frac{1}{\kappa}\left[A^2\left\{\frac{1}{A^2}\left(\frac{f-Rf_R}{2}
-\frac{\dot{f_R}}{A^2}\left(\frac{\dot{B}}{B}+\frac{2\dot{C}}{C}\right)-\frac{f'_R}{B^2}\left(\frac{B'}{B}
-\frac{2C'}{C}\right)\right.\right.\right.\\\nonumber
&&\left.\left.\left.+\frac{f''_R}{B^2}\right)\right\}_{,0}
+A^2\left\{\frac{1}{A^2B^2}\left(\dot{f'_R}-\frac{A'}{A}\dot{f_R}
-\frac{\dot{B}}{B}f_R'\right)\right\}_{,1}-\frac{\dot{f_R}}{A^2}\left\{\left(\frac{3A'}{A}\right.\right.\right.\\\nonumber
&&\left.\left.\left.+\frac{B'}{B}+\frac{2C'}{C}\right)\frac{AA'}{B^2}
+\left(\frac{\dot{B}}{B}\right)^2+2\left(\frac{\dot{C}}{C}\right)^2
+\frac{3\dot{A}}{A}\left(\frac{\dot{B}}{B}+\frac{2\dot{C}}{C}\right)
\right\}\right.\\\nonumber
&&\left.+\frac{\dot{f'_R}}{B^2}\left(\frac{3A'}{A}
+\frac{B'}{B}+\frac{2C'}{C}\right)-
\frac{2f'_R}{B^2}\left\{\frac{A'}{A}\left(\frac{2\dot{B}}{B}+\frac{\dot{C}}{C}\right)+\frac{B'}{B}\left(\frac{\dot{A}}{A}
\right.\right.\right.\\\nonumber
&&\left.\left.\left.+\frac{\dot{B}}{B}\right)-
\frac{C'}{C}\left(\frac{2\dot{A}}{A}-\frac{\dot{B}}{B}+\frac{\dot{C}}{C}\right)\right\}
+\frac{f_R''}{B^2}\left(\frac{2\dot{A}}{A}+\frac{\dot{B}}{B}\right)+\frac{\dot{A}}{A}(f\right.\\\label{B3}
&&\left.
-Rf_R)+\frac{\ddot{f_R}}{A^2}\left(\frac{\dot{B}}{B}+\frac{2\dot{C}}{C}\right)\right],
\\\nonumber
P_2(r,t)&=&\frac{1}{\kappa}\left[B^2\left\{\frac{1}{B^2}\left(\frac{Rf_R-f}{2}
-\frac{\dot{f_R}}{A^2}\left(\frac{\dot{A}}{A}-\frac{2\dot{C}}{C}\right)-\frac{f'_R}{B^2}\left(\frac{A'}{A}
+\frac{2C'}{C}\right)\right.\right.\right.
\\\nonumber &&\left.\left.\left.+\frac{\ddot{f_R}}{A^2}\right)\right\}_{,1}
+B^2\left\{\frac{1}{A^2B^2}\left(\dot{f_R}'-\frac{A'}{A}\dot{f_R}
-\frac{\dot{B}}{B}f_R'\right)\right\}_{,0}+\frac{A'}{A}\left\{\frac{\ddot{f_R}}{A^2}
\right.\right.
\\\nonumber
&&\left.\left.+\frac{f_R''}{B^2}-\frac{\dot{f_R}}{A^2}\left(\frac{\dot{A}}{A}+\frac{\dot{B}}{B}\right)-
\frac{f'_R}{B^2}\left(\frac{A'}{A} +\frac{B'}{B}\right)\right\}
+\frac{2B'}{B}\left\{\frac{Rf_R-f}{2}\right.\right.\\\nonumber
&&\left.\left.+\frac{\ddot{f_R}}{A^2}
-\frac{\dot{f_R}}{A^2}\left(\frac{\dot{A}}{A}-\frac{2\dot{C}}{C}\right)
-\frac{f'_R}{B^2}\left(\frac{A'}{A}+\frac{3C'}{C}\right)\right\}
+\frac{1}{A^2}\left(\frac{\dot{A}}{A}+\frac{3\dot{B}}{B}\right.\right.
\\\nonumber &&\left.\left.
+\frac{2\dot{C}}{C}\right)\left(\dot{f_R}'-\frac{A'}{A}\dot{f_R}
-\frac{\dot{B}}{B}f_R'\right)+\frac{2C'}{C}\left\{\frac{f''_R}{B^2}+\frac{\dot{f_R}}{A^2}\left(\frac{\dot{C}}{C}
-\frac{2\dot{B}}{B}\right)\right.\right.\\\label{B4} &&\left.\left.
-\frac{f'_R}{B^2}\frac{C'}{C}\right\}\right].
\end{eqnarray}

\section*{Appendix B: Perturbed Quantities}

\begin{eqnarray}\nonumber
P_{1p}&=&\frac{1}{\kappa}\left[-\left\{\frac{1}{A_0^2B_0^2}\left[\alpha
n(n-1)R_0^{n-2}\left(e'+((n-2)R_0^{-1}R_0'-\frac{A_0'}{A_0})e\right.\right.\right.\right.\\\nonumber
&&\left.\left.\left.\left.-
\frac{bR_0'}{B_0}\right)+\mu^4m(m+1)R_0^{-m-2}\left(-e'+(m+2)R_0^{-1}R_0'e+\frac{eA_0'}{A_0}
\right.\right.\right.\right.\\\nonumber
&&\left.\left.\left.\left.+\frac{bR_0'}{B_0}\right)\right]\right\}'+\left\{-\frac{1}{A_0^2B_0^2}\left[\alpha(n-1)B_0^2R_0^n\left(\frac{n
R_0^{-1}e}{2}+\frac{a}{A_0}\right) \right.\right.\right.\\\nonumber
&&\left.\left.\left.+\mu^4(m+1)B_0^2R_0^{-m}\alpha\left(\frac{-mR_0^{-1}}{2}-\frac{a}{A_0}\right)
-n(n-1)R_0^{n-2}\left\{e''\right.\right.\right.\right.\\\nonumber
&&\left.\left.\left.+2(n-2)R_0^{-1}R_0'e'-2\left(\frac{a}{A_0}+
\frac{b}{B_0}\right)\left( (n-2)R_0^{-1} R_0'^2+R_0''
\right)\right.\right.\right.\\\nonumber
&&\left.\left.\left.+(n-2)(n-3)R_0^{-2}R_0'^2e+(n-2)R_0^{-1}R_0''e-\left[\left(\frac{b}{B_0}\right)'
-2\left(\frac{\bar{c}}{r}\right)'\right.\right.\right.\right.\\\nonumber
&&\left.\left.\left.\left.\left.-2\left(\frac{a}{A_0}+
\frac{b}{B_0}\right)\left(\frac{B_0'}{B_0}-\frac{2}{r}\right)\right]R_0'-\left(\frac{B_0'}{B_0}-\frac{2}{r}\right)\left(
(n-2)R_0^{-1}R_0'e\right.\right.\right.\right.\right.\\\nonumber
&&\left.\left.\left.\left.\left.+e'\right)\right\}-\mu^4
m(m+1)R_0^{-m-2}\left(2(m+2)R_0^{-1}R_0'e'+(m+2)R_0^{-1}R_0''e
\right.\right.\right.\right.\\\nonumber &&\left.\left.\left.\left.
-e''-(m+2)(m+3)R_0^{-2}R_0'^2e-2\left(\frac{a}{A_0}+
\frac{b}{B_0}\right)\left((m+2)R_0^{-1}R_0'^2\right.\right.\right.\right.\right.\\\nonumber
&&\left.\left.\left.\left.\left.-R_0''\right)-\left[
2\left(\frac{a}{A_0}+
\frac{b}{B_0}\right)\left(\frac{B_0'}{B_0}-\frac{2}{r}\right)-\left(\frac{b}{B_0}\right)'+2\left(\frac{\bar{c}}{r}\right)'\right]R_0'
+\left((m\right.\right.\right.\right.\right.\\\nonumber
&&\left.\left.\left.\left.\left.+2)R_0^{-1}R_0'e\right)\left(\frac{B_0'}{B_0}-
\frac{2}{r}\right)-e'\right)\right]\right\}_{,0}\alpha(n-1)R_0^n\left\{\frac{2a}{A_0^3B_0^2}
\left(-\frac{B_0^2}{2}+\right.\right.\right.\\\nonumber
&&\left.\left.\left.(n-1)\left[(n-2)R_0^{-1}R_0'^{2}+R_0''\right]+nR_0^{-2}R_0'
\left(\frac{B_0'}{B_0}-\frac{2}{r}\right)\right)+
\left(R_0''+(n\right.\right.\right.\\\nonumber
&&\left.\left.\left.-2)R_0^{-1}R_0'^2-R_0'\left(\frac{A_0'}{A_0}+\frac{B_0'}{B_0}\right)\right)\frac{bnR_0^{-2}}{A_0^2B_0^3}
+\frac{2n\bar{c}R_0^{-2}R_0'}{rA_0^2B_0^2}\left(\frac{1}{r}-\frac{A_0'}{A_0}\right)
-\right.\right.\\\nonumber
&&\left.\left.\frac{nR_0^{-2}}{A_0^2B_0^2}\left(e'+
(n-2)R_0^{-1}R_0'e-\frac{eA_0'}{A_0}-\frac{bR_0'}{B_0}\right)+\left(\frac{3A_0'}{A_0}+\frac{B_0'}{B_0}+\frac{2}{r}\right)\right\}
\right.\\\nonumber &&\left.
+\mu^4(m+1)R_0^{-m}\left\{\frac{2a}{A_0^3B_0^2}\left(mR_0^{-2}R_0'\left(\frac{B_0'}{B_0}-\frac{2}{r}\right)
+\frac{B_0^2}{2}+(m+1)\left[R_0''\right.\right.\right.\right.\\\nonumber
&&\left.\left.\left.\left.-(m+2)R_0^{-1}R_0'^{2}\right]\right)
+\frac{bmR_0^{-2}}{A_0^2B_0^3}\left((m+2)R_0^{-1}R_0'^2+
R_0'\left(\frac{A_0'}{A_0}+\frac{B_0'}{B_0}\right)\right.\right.\right.\\\nonumber
&&\left.\left.\left.
-R_0''\right)-\frac{mR_0^{-2}}{A_0^2B_0^2}\left(-e'
-\frac{2m\bar{c}R_0^{-2}R_0'}{rA_0^2B_0^2}\left(\frac{1}{r}-\frac{A_0'}{A_0}\right)
+\frac{eA_0'}{A_0}+\frac{bR_0'}{B_0}\right.\right.\right.
\end{eqnarray}
\begin{eqnarray}\label{B5}
&&\left.\left.\left.+
(m+2)R_0^{-1}R_0'e\right)\left(\frac{3A_0'}{A_0}+\frac{B_0'}{B_0}+\frac{2}{r}\right)\right\}\right],
\\\nonumber
P_{2p}&=&-\frac{1}{\kappa}\left[\frac{\ddot{T}}{T}\left\{\frac{1}{A_0^2B_0^2}\left[\alpha
n(n-1)R_0^{n-2}\left(e'+(n-2)R_0^{-1}R_0'e-\frac{eA_0'}{A_0}\right.\right.\right.\right.\\\nonumber
&&\left.\left.\left.\left.-\frac{bR_0'}{B_0}\right)-\mu^4m(m+1)R_0^{-m-2}\left(e'-(m+2)R_0^{-1}R_0'e-\frac{eA_0'}{A_0}
\right.\right.\right.\right.\\\nonumber
&&\left.\left.\left.\left.-\frac{bR_0'}{B_0}\right)\right]\right\}-\left\{\frac{e}{A_0^2B_0^2}\left[\alpha
n(n-1)R_0^{n-2}-\mu^4
m(m+1)R_0^{-m-2}\right]\frac{\ddot{T}}{T}\right.\right.\\\nonumber
&&\left.\left.+\alpha
(n-1)R_0^n\left[\left[\left(\frac{a}{A_0}\right)'
+2\left(\frac{\bar{c}}{r}\right)'-\frac{4b}{B_0}\left(\frac{A_0'}{A_0}
+\frac{2}{r}\right)\right]R_0'R_0^{-2}\right.\right.\right.\\\nonumber
&&\left.\left.\left.-nR_0^{-1}e-\frac{2b}{B_0}+\left(\frac{A_0'}{A_0}+\frac{2}{r}\right)\left(e'+(n-2)R_0^{-1}R_0'e\right)R_0^{-2}\right]+\mu^4
(m\right.\right.\\\nonumber
&&\left.\left.\left.+1)R_0^{-m}\left(mR_0^{-1}e-\left[\left(\frac{a}{A_0}\right)'
+2\left(\frac{\bar{c}}{r}\right)'-\frac{4b}{B_0}\left(\frac{A_0'}{A_0}
+\frac{2}{r}\right)\right]R_0'R_0^{-2}
\right.\right.\right.\right.\\\nonumber &&\left.\left.\left.\left.
-\frac{2b}{B_0}+\left(\frac{A_0'}{A_0}+\frac{2}{r}\right)
\left(e'(n-2)R_0^{-1}R_0'e\right)R_0^{-2}\right)\right\}'+\left(\alpha
n(n-1)R_0^{n-2}\right.\right.\right.\\\nonumber
&&\left.\left.\left.-\mu^4m(m+1)R_0^{-m-2}\right)\left(\frac{2eB_0'}{A_0^2B_0^3}
-\frac{eA_0'}{A_0^3B_0^2}\right)\frac{\ddot{T}}{T}+\alpha
n(n-1)R_0^{n-2}\left(e''\right.\right.\right.\\\nonumber
&&\left.\left.+2(n-2)R_0^{-1}R_0'e'+(n-2)(n-3)R_0^{-2}R_0'^2e+(n-2)R_0^{-1}R_0''e\right.\right.\\\nonumber
&&\left.\left.\left.-\frac{2b}{B_0}\left(R_0''+(n-2)R_0^{-1}R_0'^2\right)\right)\left(-\frac{A_0'}{A_0B_0}-\frac{2}{rB_0^2}\right)+\alpha
n(n-1)R_0^{n-2}\right.\right.\\\nonumber &&\times\left.\left.
\left(\left[\left(-\frac{A_0'}{A_0B_0}-\frac{2B_0'}{B_0}\right)\left(\frac{a}{A_0}\right)'
+\left(-\frac{A_0'}{A_0B_0}+
\frac{2}{rB_0^2}\right)\left(\frac{b}{B_0}\right)'
\right.\right.\right.\right.\\\nonumber
&&\left.\left.\left.+\left(\frac{4B_0'}{B_0}+\frac{2}{rB_0^4}\right)\left(\frac{\bar{c}}{r}\right)'
+\frac{2A_0'}{A_0B_0^2}\left(\frac{A_0'}{A_0}+\frac{B_0'}{B_0}-\frac{4bB_0'}{r^2}\right)+\frac{8bB_0'}{rB_0^6}\right]R_0'
\right.\right.\\\nonumber
&&\left.\left.+(e'+(n-2)R_0^{-1}R_0'e')\left[-\frac{A_0'}{A_0B_0}\left(\frac{A_0'}{A_0}
+\frac{B_0'}{B_0}-2B_0'\right)
+\frac{2}{rB_0}\left(\frac{B_0'}{B_0^4}
\right.\right.\right.\right.\\\nonumber
&&\left.\left.\left.\left.+\frac{1}{rB_0^3}+2B_0'\right)\right]-\alpha
n(n-1)R_0^{n-2}\left[\frac{A_0'}{A_0B_0^2}\left((n-2)R_0^{-1}R_0'^2+R_0''\right.\right.\right.\right.\\\nonumber
&&\left.\left.\left.\left.+\left(\frac{A_0'}{A_0}+\frac{B_0'}{B_0}
\right)R_0'\right)\left(\frac{a'}{A_0'}-\frac{a}{A_0}-\frac{2b}{B_0}\right)
+\frac{2}{rB_0^4}\left((n-2)R_0^{-1}R_0'^2\right.\right.\right.\right.
\end{eqnarray}
\begin{eqnarray}\nonumber
&&\left.\left.\left.\left.+R_0''+\left(\frac{B_0'}{B_0}+
\frac{1}{r}\right)R_0'\right)\left(\bar{c}-\frac{1}{r}-\frac{2b}{B_0}\right)\right]
+\alpha(n-1)R_0^n\left[\frac{2b}{B_0}
\right.\right.\right.\\\nonumber&&\left.\left.\left.-nR_0^{-1}e+
\frac{1}{B_0^2}\left(\frac{b'}{B_0'}-\frac{b}{B_0}\right)\left(-\frac{1}{2}+\frac{nR_0^{-2}R_0'}{B_0^2}\left(\frac{A_0'}{A_0}+
\frac{2}{r}\right)\right)\right]\frac{2B_0'}{B_0}
\right.\right.\\\nonumber &&\left.\left.+\mu^4
m(m+1)R_0^{-m-2}\left(-\frac{A_0'}{A_0B_0}-\frac{2}{rB_0^2}\right)\left(-e''+2(m+2)R_0^{-1}R_0'e
\right.\right.\right.
\\\nonumber
&&\left.\left.-(m+2)(m+3)R_0^{-2}R_0'^2e+(m+2)R_0^{-1}R_0''e
-\frac{2b}{B_0}\left((m+2)R_0^{-1}R_0'^2\right.\right.\right.\\\nonumber
&&\left.\left.\left.\left.-R_0''\right)\right)+\mu^4
m(m+1)R_0^{-m-2}\left(-\left[\left(-\frac{A_0'}{A_0B_0}-\frac{2B_0'}{B_0}\right)\left(\frac{a}{A_0}\right)'
\right.\right.\right.\right.\\\nonumber
&&\left.\left.\left.\left.+\left(-\frac{A_0'}{A_0B_0}+
\frac{2}{rB_0^2}\right)\left(\frac{b}{B_0}\right)'+\left(\frac{4B_0'}{B_0}+\frac{2}{rB_0^4}\right)\left(\frac{\bar{c}}{r}\right)'
+\frac{2A_0'}{A_0B_0^2}\left(\frac{A_0'}{A_0}\right.\right.\right.\right.\right.\\\nonumber
&&\left.\left.\left.\left.+\frac{B_0'}{B_0}-\frac{4bB_0'}{r^2}\right)+\frac{8bB_0'}{rB_0^6}\right]R_0'+(-e'+(m+2)R_0^{-1}R_0'e')
\left[\frac{2}{rB_0}\left(\frac{B_0'}{B_0^4}\right.\right.\right.\right.\\\nonumber
&&\left.\left.\left.\left.
+\frac{1}{rB_0^3}+2B_0'\right)-\frac{A_0'}{A_0B_0}\left(\frac{A_0'}{A_0}+\frac{B_0'}{B_0}-2B_0'\right)
\right]-\mu^4 m(m+1)R_0^{-m-2} \right.\right.\\\nonumber
&&\times\left.\left.\left[\frac{A_0'}{A_0B_0^3}\left(-(m+2)R_0^{-1}R_0'^2+R_0''+\left(\frac{A_0'}{A_0}+
\frac{B_0'}{B_0}\right)R_0'\right)\left(\frac{a'}{A_0'}-\frac{a}{A_0}
\right.\right.\right.\right.\\\nonumber &&\left.\left.\left.\left.
-\frac{2b}{B_0}\right)
+\frac{2}{rB_0^4}\left(R_0''-(m+2)R_0^{-1}R_0'^2+\left(\frac{B_0'}{B_0}+
\frac{1}{r}\right)R_0'\right)+\left(\bar{c}-\frac{1}{r}\right.\right.\right.\right.\\\nonumber
&&\left.\left.\left.-\frac{2b}{B_0}\right)\right]+\mu^4
(m+1)R_0^{-m}\left[\frac{1}{B_0^2}\left(\frac{b'}{B_0'}-\frac{b}{B_0}\right)\left(
\frac{mR_0^{-2}R_0'}{B_0^2}\left(\frac{A_0'}{A_0}+
\frac{2}{r}\right)
\right.\right.\right.\\\label{B6}&&\left.\left.\left.+\frac{1}{2}\right)
+(mR_0^{-1}e-\frac{2b}{B_0})\right]\frac{2B_0'}{B_0}\right]T,
\\\nonumber
P_3&=&2\left(\frac{A_0'}{B_0}+\frac{A_0}{rB_0}\right)\left[\frac{\alpha
R_0^{n-1}}{\kappa
A_0B_0}\left\{(n-1)R_0^{-1}+(n-1)(n-2)R_0^{-2}R_0'e\right.\right.\\\nonumber
&&\left.\left.\left.-(n-1)R_0^{-1}\left(\frac{eA_0'}{A_0}+\frac{b}{B_0}\right)
+2\left(\frac{\bar{c}'}{r}-\frac{\bar{c}A_0'}{rA_0}-\frac{b}{rB_0}\right)\right\}
+\frac{\mu^4 m R_0^{-m-1}}{\kappa A_0B_0}\right.\right.\\\nonumber
&&\times\left.\left.\{-(m+1)R_0^{-1}e'+(m+1)(m+2)R_0^{-2}R_0'e+(m+1)R_0^{-1}\left(\frac{eA_0'}{A_0}
\right.\right.\right.\\\nonumber
&&\left.\left.\left.+\frac{b}{B_0}\right)+2\left(\frac{\bar{c}'}{r}-\frac{\bar{c}A_0'}{rA_0}-\frac{b}{rB_0}\right)\right\}+\frac{2}{\kappa
A_0B_0}\left(\frac{\bar{c}'}{r}-\frac{\bar{c}A_0'}{rA_0}-\frac{b}{rB_0}\right)\right]\dot{T}\\\nonumber
&&+\frac{A_0}{B_0}\left[\frac{\alpha R_0^{n-1}}{\kappa
A_0B_0}\left\{(n-1)R_0^{-1}+(n-1)(n-2)R_0^{-2}R_0'e-(n-1)R_0^{-1}\right.\right.
\end{eqnarray}
\begin{eqnarray}
\nonumber
&&\times\left.\left.\left.\left(\frac{eA_0'}{A_0}+\frac{b}{B_0}\right)+2\left(\frac{\bar{c}'}{r}-\frac{\bar{c}A_0'}{rA_0}-\frac{b}{rB_0}\right)\right\}
+\frac{\mu^4 m R_0^{-m-1}}{\kappa
A_0B_0}\{-R_0^{-1}e'(m\right.\right.\\\nonumber
&&\left.\left.+1)+(m+1)(m+2)R_0^{-2}R_0'e+(m+1)R_0^{-1}\left(\frac{eA_0'}{A_0}+\frac{b}{B_0}\right)
+2\left(\frac{\bar{c}'}{r} \right.\right.\right.\\\label{B7}
&&\left.\left.\left.-\frac{\bar{c}A_0'}{rA_0}-\frac{b}{rB_0}\right)\right\}+\frac{2}{\kappa
A_0B_0}\left(\frac{\bar{c}'}{r}-\frac{\bar{c}A_0'}{rA_0}-\frac{b}{rB_0}\right)\right]'\dot{T}+A_0^2
P_{1p},
\\\nonumber
P_{4}&=&-\frac{rA_0^2B_0}{br+2B_0\bar{c}}\left[
\frac{e}{2}-\frac{2\bar{c}}{r^3}
-\frac{1}{A_0B_0^2}\left\{A_0''[\frac{a}{A_0}+\frac{2b}{B_0}]-\frac{1}{B_0}\left(a'B_0'+A_0'b'
\right.\right.\right.\\\nonumber && \left.\left.\left.
+a''-A_0'B_0'[\frac{a}{A_0}+\frac{3b}{B_0}]\right)
+\frac{2}{r}\left\{a'+\bar{c}'A_0'-A_0'[\frac{a}{A_0}+\frac{2b}{B_0}+\frac{\bar{c}}{r}]\right\}
\right.\right.\\\label{B9}&& \left.\left.
+\frac{A_0}{r}\left\{\bar{c}''-\frac{b'}{B_0}-\frac{B_0'\bar{c}'}{B_0}+
\frac{3b}{B_0}+\frac{\bar{c}}{r}\right\}
+\frac{2}{r^2}[\bar{c}'-\frac{b}{B_0}-\frac{\bar{c}}{r}]\right\}\right],
\\\nonumber
P_{2P}^{N}&=&-\frac{1}{\kappa}\left[\left\{\alpha
n(n-1)R_0^{n-2}\left(e'+(n-2)R_0^{-1}R_0'e-\mu^4m(m+1)R_0^{-m-2}
\right.\right.\right.\\\nonumber &&\left.\left.\left.
-bR_0'\right)\left(e'-(m+2)R_0^{-1}R_0'e-bR_0'\right)\right\}\frac{\ddot{T}}{T}+\left\{e\left[\alpha
n(n-1)R_0^{n-2}\right.\right.\right.\\\nonumber
&&\left.\left.\left.-\mu^4
m(m+1)R_0^{-m-2}\right]\frac{\ddot{T}}{T}+\alpha
(n-1)R_0^n\left[-nR_0^{-1}e-2b+\left[2\left(\frac{\bar{c}}{r}\right)'\right.\right.\right.\right.\\\nonumber
&&\left.\left.\left.\left. +a'-\frac{8b}{r}
\right]+\frac{2}{r}\left(e'(n-2)R_0^{-1}R_0'e\right)R_0^{-2}\right]R_0'R_0^{-2}+\mu^4
(m+1)R_0^{-m}\right.\right.\\\nonumber
&&\left.\times\left(mR_0^{-1}e-2b
-\left[a+2\left(\frac{\bar{c}}{r}\right)'-\frac{8b}{r}\right]R_0'R_0^{-2}+\frac{2}{r}\right)
\left(R_0^{-2}+e'(n \right.\right.\\\nonumber
&&\left.\left.\left.\left. -2)R_0^{-1}R_0'e\right)\right\}'-\alpha
n(n-1)R_0^{n-2}\left(-\frac{2}{r}\right)\left(e''+2(n-2)R_0^{-1}R_0'e'\right.\right.\right.\\\nonumber
&&\left.\left.\left.+(n-2)(n-3)R_0^{-2}R_0'^2e+(n-2)R_0^{-1}R_0''e-2b\left((n-2)R_0^{-1}R_0'^2
\right.\right.\right.\right.\\\nonumber
&&\left.\left.\left.\left.+R_0''\right)\right)+\alpha
n(n-1)R_0^{n-2}\left(\left[\left(
\frac{2}{r}\right)b'+\frac{2}{r}\left(\frac{\bar{c}}{r}\right)'
\right]R_0'+\frac{2}{r^2}(e'+(n \right.\right.\right.\\\nonumber
&&\left.\left.-2)R_0^{-1}R_0'e')\right) -\alpha
n(n-1)R_0^{n-2}\left[\frac{2}{r}\left(\bar{c}-\frac{1}{r}-2b\right)\left(R_0''
+\frac{1}{r}R_0'\right.\right.\right.\\\nonumber
&&\left.\left.\left.\left.+(n-2)R_0^{-1}R_0'^2\right)\right] +\mu^4
m(m+1)R_0^{-m-2}\left(-\frac{2}{r}\right)\left(2(m+2)R_0^{-1}R_0'e\right.\right.\right.\\\nonumber
&&\left.\left.\left.-e''
-(m+2)(m+3)R_0^{-2}R_0'^2e-2b\left((m+2)R_0^{-1}R_0'^2-R_0''\right)+(m
\right.\right.\right.\\\nonumber
&&\left.\left.+2)R_0^{-1}R_0''e\right)+\mu^4
m(m+1)R_0^{-m-2}\left(-\left[\frac{2b'}{r}+\frac{2}{r}\left(\frac{\bar{c}}{r}\right)'
\right]R_0'+(-e'\right.\right.
\end{eqnarray}
\begin{eqnarray}\nonumber
&&\left.\left.+(m+2)R_0^{-1}R_0'e')\left(\frac{2}{r^2}\right)-\mu^4
m(m+1)R_0^{-m-2}\left(R_0''+ \frac{1}{r}R_0'
\right.\right.\right.\\\label{B2N}
&&\left.\left.\left.-(m+2)R_0^{-1}R_0'^2\right)\left(\bar{c}-\frac{1}{r}-2b\right)\left(\frac{2}{r}\right)\right]
T,\right.\\\nonumber P_5&=&P_4\left[\frac{\alpha n R_0^{n-1}}{\kappa
}\left[(n-1)R_0^{-1}e'+(n-1)(n-2)R_0^{-2}R_0'e+2\left(\frac{\bar{c}'}{r}\right.\right.\right.\\\nonumber
&&\left.\left.\left.-\frac{b}{r}\right)-(n-1)R_0'b\right]+\frac{\mu^4
m R_0^{-m-1}}{\kappa
}\left[-(m+1)R_0^{-1}e'+(m+1)\right.\right.\\\label{P5}
&&\left.\left.(m+2)R_0^{-2}R_0'e+(m+1)R_0^{-1}b\right]+\frac{2}{\kappa
}\left(\frac{\bar{c}'}{r}-\frac{b}{r}\right)\right]+P_{2P}^{N}.
\end{eqnarray}

\end{document}